\documentclass[fleqn,usenatbib]{mnras}

\usepackage{newtxtext,newtxmath}
\usepackage[T1]{fontenc}

\DeclareRobustCommand{\VAN}[3]{#2}
\let\VANthebibliography\thebibliography
\def\thebibliography{\DeclareRobustCommand{\VAN}[3]{##3}\VANthebibliography}

\usepackage{graphicx}	% Including figure files
\usepackage{amsmath}	% Advanced maths commands
\usepackage{layouts}
\usepackage{siunitx} % Displays units correctly
\newcommand{\tavg}[1]{\langle #1 \rangle}
\newcommand{\scarequotes}[1]{`#1'}
\newcommand{\dn}[1]{\,\mathrm{d}#1}

\newcommand{\Ks}{\ensuremath{K_\mathrm{s}}}

\DeclareSIUnit{\erg}{erg}
\DeclareSIUnit{\parsec}{pc}
\newcommand{\dmunits}{\ensuremath{\ \mathrm{pc}\ \mathrm{cm}^{-3}}}
\DeclareSIUnit{\pccm}{\parsec\ \centi\meter}
\DeclareSIUnit{\dmunits}{\pccm^{-3}}

\newcommand{\m}[1]{\ensuremath{m_\mathrm{#1}}}

\newcommand{\DM}[1]{\ensuremath{\mathrm{DM_{#1}}}}
\newcommand{\DMFRB}{\DM{FRB}}
\newcommand{\DMHost}{\DM{host}}
\newcommand{\DMHostISM}{\DM{host,ISM}}
\newcommand{\DMCosmic}{\DM{cosmic}}
\newcommand{\DMExgal}{\DM{exgal}}
\newcommand{\DMMW}{\DM{MW}}
\newcommand{\DMMWHalo}{\DM{MW,halo}}
\newcommand{\DMMWISM}{\DM{MW,ISM}}

\newcommand{\frb}{_\mathrm{FRB}}
\newcommand{\tauFRB}{\ensuremath{\tau\frb}}
\newcommand{\widthintFRB}{\sigma\frb}
\newcommand{\zhost}{\ensuremath{z_\mathrm{host}}}

\newcommand{\pzdm}{\ensuremath{p(z|{\rm DM})}}

\newcommand{\rarl}[4]{$#1^\mathrm{h}#2^\mathrm{m}#3\fs#4$}
\newcommand{\decrl}[4]{$#1 \si{\degree} #2 \si{\arcminute} #3 \farcs #4$}

\newcommand{\FRB}[1]{FRB\,#1}

\newcommand{\FRBname}{\FRB{20210912A}} 
\newcommand{\FRBDM}{1233.696}
\newcommand{\FRBDMerr}{0.006}
\newcommand{\FRBSN}{31.7}
\newcommand{\FRBfreq}{1271.5} % MHz
\newcommand\FRBDMMWISMNE{31}
\newcommand\FRBDMMWISMYMW{17}
\newcommand\FRBDMMWHalo{50}
\newcommand\FRBDMEG{1153.6}

\newcommand{\FRBRA}{\rarl{23}{23}{10}{35}}
\newcommand{\FRBDec}{\decrl{-30}{24}{19}{2}}
\newcommand{\FRBwidthtotal}{0.085} % milliseconds
\newcommand\FRBtau{59\pm4} % microseconds
\newcommand\FRBwidth{26\pm2} % micros econds
\newcommand\FRBfreqtau{1263.5} %MHz
\newcommand{\raerr}{0.4 arcsec}
\newcommand{\decerr}{0.3 arcsec}
\newcommand{\FRBfluence}{69.9} %Jy ms

\newcommand{\FRBnameB}{\FRB{20210407A}}
\newcommand{\gmaglim}{26.3}
\newcommand{\Rmaglim}{26.7}
\newcommand{\Imaglim}{24.2}
\newcommand{\Kmaglim}{24.9}

\newcommand{\gext}{0.05}
\newcommand{\Rext}{0.03}
\newcommand{\Iext}{0.02}
\newcommand{\Kext}{0.004}

\newcommand{\gfwhm}{1.0}
\newcommand{\Rfwhm}{1.0}
\newcommand{\Ifwhm}{1.0}
\newcommand{\Kfwhm}{0.3}

\newcommand{\PATHradius}{11\,arcsec}

\newcommand{\NCRAFTHosts}{25}

\title[Unseen FRB host]{The unseen host galaxy and high dispersion measure of a precisely-localised Fast Radio Burst suggests a high-redshift origin}

\author[L. Marnoch et al.]{Lachlan~Marnoch,$^{1,2,3,4}$\thanks{E-mail: lachlan.marnoch@hdr.mq.edu.au}
Stuart~D.~Ryder,$^{1,3}$
Clancy~W.~James,$^5$ 
Alexa~C.~Gordon,$^6$ 
Mawson~W.~Sammons,$^{5}$
\newauthor
J.~Xavier~Prochaska,$^{7,8,9}$ 
Nicolas~Tejos$^{10}$ 
Adam~T.~Deller,$^{11}$ 
Danica~R.~Scott,$^{5}$ 
Shivani~Bhandari,$^{2,12,13,14}$ 
\newauthor
Marcin~Glowacki,$^{5}$ 
Elizabeth~K.~Mahony,$^{2}$
Richard~M.~McDermid,$^{1,3,4}$
Elaine M. Sadler$^{15,2,4}$
\newauthor
Ryan~M.~Shannon,$^{11}$ 
and Hao~Qiu,$^{16}$
\\
$^{1}$School of Mathematical and Physical Sciences, Macquarie University, Sydney, NSW 2109, Australia\\
$^{2}$Australia Telescope National Facility, CSIRO Space and Astronomy, P.O. Box 76, Epping, NSW 1710, Australia\\
$^{3}$Astrophysics and Space Technologies Research Centre, Macquarie University, Sydney, NSW 2109, Australia\\
$^{4}$ARC Centre of Excellence for All-Sky Astrophysics in 3 Dimensions (ASTRO 3D), Australia\\
$^{5}$International Centre for Radio Astronomy Research, Curtin University, Bentley, WA 6102, Australia\\
$^{6}$Center for Interdisciplinary Exploration and Research in Astrophysics (CIERA) and Department of Physics and Astronomy, \\Northwestern University, Evanston, IL 60208, USA\\
$^{7}$Department of Astronomy and Astrophysics, University of California, Santa Cruz, CA 95064, USA\\
$^{8}$Kavli Institute for the Physics and Mathematics of the Universe (Kavli IPMU), 5-1-5 Kashiwanoha, Kashiwa, 277-8583, Japan\\
$^{9}$Division of Science, National Astronomical Observatory of Japan,2-21-1 Osawa, Mitaka, Tokyo 181-8588, Japan\\
$^{10}$Instituto de F\'isica, Pontificia Universidad Cat\'olica de Valpara\'iso, Casilla 4059, Valpara\'iso, Chile\\
$^{11}$Centre for Astrophysics and Supercomputing, Swinburne University of Technology, Hawthorn, VIC 3122, Australia\\
$^{12}$ASTRON, Netherlands Institute for Radio Astronomy, Oude Hoogeveensedijk 4, 7991 PD Dwingeloo, The Netherlands\\
$^{13}$Joint institute for VLBI ERIC, Oude Hoogeveensedijk 4, 7991 PD Dwingeloo, The Netherlands\\
$^{14}$Anton Pannekoek Institute for Astronomy, University of Amsterdam, Science Park 904, 1098 XH, Amsterdam, The Netherlands\\
$^{15}$Sydney Institute for Astronomy, School of Physics A28, University of Sydney, NSW 2006, Australia\\
$^{16}$SKA Observatory, SKA Observatory, Jodrell Bank, Lower Withington, Macclesfield, SK11 9FT, UK\\
}

\date{Accepted 2023 July 31. Received 2023 July 27; in original form 2023 June 18}

\pubyear{2023}

\begin{document}
\label{firstpage}
\pagerange{\pageref{firstpage}--\pageref{lastpage}}
\maketitle

\begin{abstract}
FRB\,20210912A is a fast radio burst (FRB), detected and localised to sub-arcsecond precision by the Australian Square Kilometre Array Pathfinder. No host galaxy has been identified for this burst despite the high precision of its localisation and deep optical and infrared follow-up, to 5-$\sigma$ limits of $R=\Rmaglim$\,mag and $\Ks=\Kmaglim$\,mag with the Very Large Telescope. The combination of precise radio localisation and deep optical imaging has almost always resulted in the secure identification of a host galaxy, and this is the first case in which the line-of-sight is not obscured by the Galactic disk. The dispersion measure of this burst, $\DMFRB{}=\FRBDM{}\pm\FRBDMerr{}\dmunits$, allows for a large source redshift of $z>1$ according to the Macquart relation. It could thus be that the host galaxy is consistent with the known population of FRB hosts, but is too distant to detect in our observations ($z>0.7$ for a host like that of the first repeating FRB source, \FRB{20121102A}); that it is more nearby with a significant excess in $\DM{host}$, and thus dimmer than any known FRB host; or, least likely, that the FRB is truly hostless. We consider each possibility, making use of the population of known FRB hosts to frame each scenario.
The fact of the missing host has ramifications for the FRB field: even with high-precision localisation and deep follow-up, some FRB hosts may be difficult to detect, with more distant hosts being the less likely to be found. This has implications for FRB cosmology, in which high-redshift detections are valuable.
\end{abstract}


% galaxies: luminosity function

%%%%%%%%%%%%%%%%%%%%%%%%%%%%%%%%%%%%%%%%%%%%%%%%%%

%%%%%%%%%%%%%%%%% BODY OF PAPER %%%%%%%%%%%%%%%%%%

\section{Introduction}

Fast radio bursts (FRBs) are
intense, short-duration pulses of radio emission, currently being searched for on the majority of the world's radio
facilities. Astronomers have been compelled to study the bursts for two overarching purposes: 
  (1) to identify the origin(s) of the enigmatic bursts
  to elucidate the astrophysical sources and mechanisms that cause them;
  and
  (2) to leverage measurements from their signals to study
  fundamental properties of our Universe.
On the latter, the dispersion measure (DM, 
the primary observable of an FRB)
provides a direct measurement of the integrated
ionised electron density along the sightline to the source. The DM can be broadly split into three components, $\DMFRB{} = \DMMW{} + \DMCosmic{} + \DMHost{}$. 
The Galactic contribution, \DMMW{}, includes the interstellar medium and the halo of the Milky Way.
\DMCosmic{} includes the contribution from the 
intergalactic medium (IGM) and intervening
galaxies. 
\DMHost{} is the contribution of the host galaxy, including the halo, interstellar medium and immediate FRB environment.
When coupled with spectroscopic redshift determinations \zhost, which are established by precisely
localising an FRB to an associated host galaxy (e.g. \citealp{HostR1, FRB180924}),
it becomes possible to map the otherwise invisible (ionised) cosmic web of the Universe.
The resultant Macquart relation between $\DMCosmic{}$ and \zhost\
reveals the previously ``missing'' baryons \citep{MacquartCosmicDM}
and offers a powerful approach to resolving the
underlying large-scale structure \citep{FLIMFLAM}. 

The Macquart relation has lately been leveraged to place unique and complementary constraints on the 
Hubble constant $H_0$ \citep{JamesH0}
and galactic feedback processes \citep{Baptista2023}, 
and the FRB field aspires to ultimately test scenarios of helium \citep{Caleb2019a} and hydrogen reionisation \citep{walters2018,Zhao2020}.
These new cosmological studies will greatly benefit 
from the detection of $z\gtrsim1$ FRBs and 
the subsequent
follow-up of their (presumably) faint host galaxies.
The benefits of higher-redshift FRBs for cosmology are twofold: first, the impact of structure in the cosmic web becomes fractionally smaller as a longer path length is traversed \citep{Baptista2023}; and second, the uncertain contribution of ionised material in the host galaxy is attenuated by a factor of $(1+z)^{-1}$, while also becoming fractionally smaller. Combined, these effects make higher-redshift FRBs superior for probing cosmological evolution.

FRB detections in this redshift regime also have implications for progenitor hypotheses.
For an FRB to be detectable from $z>1$ requires a radio energy
of typically $10^{41}-10^{42}$\,erg \citep{RyderFRB20220610A}--challenging FRB source
models and potentially nearing the theoretical breakdown limit
of electric fields at the source \citep{LuKumar2019}.
Additionally, the leading hypothesis on the FRB
progenitor is that of young magnetars \citep{MargalitMetzger2018,chime20}. A key test is the prediction that the peak in FRB activity
matches that of star formation, near $1 < z < 3$; magnetars are relatively short-lived \citep[$\lesssim10^5$ years;][]{Magnetars} and are understood to originate primarily from core-collapse supernovae, which trace recent star formation.

FRBs cannot be reliably assigned redshifts based on the radio signals alone, as they have not been found to carry identifiable spectroscopic features. For applications requiring precision, the distance to the source must be determined by associating the burst to its host galaxy 
\citep{PATH}
and retrieving a redshift via spectroscopic follow-up \citep[e.g.][]{HostR1, FRB180924}.
Burst DM has historically been used as a rough estimate of the source distance \citep{Thornton2013}, based on models of the IGM \citep[e.g.][]{Inoue2004};
however, the sightline-dependent, and largely unknown, variance in IGM density makes this imprecise. Such methods will also overestimate the redshifts of bursts with large DM excesses over the Macquart relation. An example is \FRB{20190520B} \citep{R1-twin} with a \DMFRB{} of $1204.7~\dmunits$ but a $\zhost$ of $0.241$, implying an extreme DM excess of $> 700\,\dmunits$. This excess can be contributed either by the host galaxy ($\DMHost$) or by overdensities in the IGM and intervening halos \citep[\DMCosmic{};][]{SimhaExcess}. Indeed, although \FRB{20190520B} was believed to have the highest \DMHost{} of any FRB, there is now evidence that the halo gas of two foreground clusters is instead responsible \citep{LeeFRB20190520B}. 

So, while a given $z$ dictates a minimum DM, the reverse is not true; the DM can only be used with any certainty to place an upper limit (or, with care, a probability distribution; e.g. \citealp{Lee-WaddellFRB20171020A}) on $z$. In fact, for a very high DM, a turnover in the Macquart relation caused by detection biases makes a closer origin {\em more} likely \citep{Connor2019,James2022A}.
Even setting aside cosmic inhomogeneity, there are pulsars in our own Galaxy ($z=0$) with DM up 
to 1778~\dmunits, embedded deep within the Galactic disk \citep{Eatough2013, PSRCAT}. This suggests that, lacking a host identification, an FRB with $\DM{}\sim1000\,\dmunits$ can plausibly occupy almost any redshift up to 1. However, the well-established correlation between DM and scattering timescale in Galactic pulsars \citep{Bhat2004, Cordes2016Scattering} has been used to argue for a similar correlation between the contribution to \DMFRB{} of the host galaxy interstellar medium \DMHostISM{} and the burst scattering timescale 
$\tauFRB$ \citep{Cordes2016Scattering}, under the following assumptions: that the IGM contributes minimally to scattering, backed up by theoretical work by e.g. \citet{Macquart2013}; 
and that the host ISM behaves roughly like that of the Milky Way. Therefore the observed scattering times of FRBs may constrain the value of $\DMHost{}$, and hence improve estimates of $\DMCosmic{}$ and $\zhost$ \citep{Cordes2022}. While on solid theoretical ground, these techniques are difficult to demonstrate and calibrate in practise due to the unknown properties of a host galaxy's ISM. 

Despite the lack of precision in the \DMFRB{}-\zhost{} relationship, an FRB emitted at high redshift will certainly have a high DM. 
To date, 89 FRBs have been recorded in FRBStats \citep{FRBStats} with dispersion measures 
large enough\footnote{after subtracting the NE2001 \citep{NE2001} \DMMWISM\ for the line-of-sight and a nominal $\DMMWHalo = 40 \dmunits$ \citep{ProchaskaZheng2019}, $\DMHost = 50 \dmunits / (1 + z_\DM{})$} 
to allow for $\zhost\gtrsim1$ on the Macquart relation, suggestive of a population of FRBs
at high redshift. Of these, only two have been localised to a host.
One is \FRB{20190520B} \citep{R1-twin}, located at a lower redshift and with a large excess DM.
The other is the only known FRB host  at $\zhost>1$, that of \FRB{20220610A}, with $\DMFRB=1457.624 \dmunits$ and $\zhost=1.016$ \citep{RyderFRB20220610A}.
The detection and localisation of this FRB demonstrates that a $z\geq1$ population is likely to exist, that some FRBs emitted at this redshift--when the Universe was half of its current age--can presently be detected, and that so too can some of their hosts. However, it is likely that not all such hosts will be easily identifiable, or identifiable at all. 
It is known that some FRB hosts are galaxies with relatively low luminosity; the hosts of FRBs 20121102A \citep{HostR1}, 20210117A \citep{BhandariFRB20210117A}, and 
20190520B \citep{R1-twin} have all been claimed as 
dwarf galaxies, although definitions vary \citep{GordonProspector}. 
Placed at higher redshifts, these galaxies might entirely elude detection with current facilities while still producing detectable FRBs. In addition, the population of galaxies in the Universe extends to very low luminosities \citep[$M_V > -7.7$;][]{UFDReview}, such that the faintest in the Local Group were not discovered until 2005 \citep{Willman2005b, Willman2005a}.
It is thus likely that some proportion of FRBs will originate from host galaxies that are undetected or undetectable in optical or even near-infrared imaging; establishing redshifts in these cases will therefore be difficult or impossible. 
The fraction of FRB hosts that can be 
expected to elude detection is crucial to FRB cosmology and to the allocation of follow-up resources, which will struggle to keep pace with the onslaught of precise FRB localisations expected from (among others) the CRAFT coherent, CHIME/FRB outrigger \citep{TONE}, and DSA-2000 \citep{DSA-2000} upgrades.

Although this avalanche of FRB hosts may be just over the horizon, thus far only a few tens of FRBs have been localised to their hosts \citep{GordonProspector}. Among the groups currently performing this work, and responsible for approximately half of the published sample, is the Commensal Real-time ASKAP Fast Transients \citep[CRAFT;][]{CRAFT} survey, which uses the Australian Square Kilometre Pathfinder \citep[ASKAP;][]{ASKAP} to pinpoint single bursts to arcsecond or, frequently, sub-arcsecond precision. With only two exceptions, all CRAFT FRBs with such precise localisations and deep optical follow-up have been associated with host galaxies (totalling 25 known hosts; Shannon et al, in preparation), including the most distant confirmed to date \citep[the aforementioned \FRB{20220610A};][]{RyderFRB20220610A}. The exceptions are \FRBnameB{} and \FRBname{} \citep{JamesH0}, both with $\DMFRB{}>1000\,\dmunits$. 
The sightline to \FRBnameB{}
lies at low Galactic latitude,
which likely contributes significantly to its $\DMFRB=1785.3\,\dmunits$ and to the high Galactic 
extinction ($A_\mathrm{R} \sim 2.5$) estimated along the line-of-sight, which is believed to have hindered follow-up observations.

In this work we discuss \FRBname{}.
This was a burst discovered in commensal observations with ASKAP, first reported by \citet{JamesH0}, with a localisation precision of 0.4\,arcsec. In contrast with \FRBnameB{}, it is at high Galactic latitude ($|b| = -70.4^\circ$) where the Galaxy does not contribute significantly to extinction or the high DM.
The burst dispersion measure is 
$\FRBDM{}\pm\FRBDMerr{}\dmunits$, consistent with an origin at $z>1$; it is also, as discussed above, consistent with emission at a smaller redshift but with a large \DMHost{} and/or \DMCosmic{} contribution. 

Here we describe the search for the host galaxy of \FRBname{} and discuss potential reasons for its lack of success. Observational data in radio, optical and near-infrared, and how it was processed, is summarised in \autoref{sec:obs}. Section \ref{sec:analysis} describes how these data are combined, analysed, and compared to known FRB hosts. In \autoref{sec:scenarios}, we explore in detail the scenarios that could lead to the non-detection of the host, and discuss their plausibility before going on to discuss the means by which this missing host could be detected in the future and the implications of its non-detection to future FRB host searches.

\section{Observations}
\label{sec:obs}

\subsection{Radio detection}
\label{sec:radio}

\newcommand{\colwidth}{12cm}
\begin{table*}
    \centering
    \caption{Table of observed, derived and adopted quantities for \FRBname{}.}
    \label{tab:quantities}
\begin{tabular}{lll}
Quantity & Value & Description \\
\hline 
$\alpha_\mathrm{FRB}$           & \FRBRA{} $\pm$ 0\fs02   &
Right Ascension (J2000) of best FRB position, tied to ICRS.
\\
$\delta_\mathrm{FRB}$           & \FRBDec{} $\pm$ 0\farcs3 &
Declination (J2000) of best FRB position, tied to ICRS.
\\
$F_\mathrm{FRB}$                & \FRBfluence{} Jy ms & 
Fluence of the FRB.                             
\\
$S/N$                           & $\FRBSN{}$ &
Signal-to-noise of the initial FRB detection.
\\
$\nu_\mathrm{FRB}$              & $\FRBfreq{}\ \si{\mega\hertz}$ &
Central frequency of FRB detection window.
\\
$\sigma_\mathrm{FRB}$           & $\FRBwidth{}\ \si{\micro\second}$ &
Best-fitting intrinsic width of the FRB; see \autoref{sec:radio}
\\
$\tauFRB$                       & $\FRBtau{}\ \si{\micro\second}$ &
Best-fitting scattering timescale of the FRB; see \autoref{sec:radio}.
\\
$\nu_\mathrm{\tau}$              & $\FRBfreqtau{}\ \si{\mega\hertz}$ &
Central frequency of scattering measurements.
\\
$\DMFRB$                        & $\FRBDM \pm \FRBDMerr$ \dmunits            & 
Observed structure-maximised dispersion measure of the FRB. 

\\
$\DMMWISM$                      &  \FRBDMMWISMNE \dmunits & 
Contribution of the Galactic ISM to $\DMFRB$ according to the NE2001 model \citep{NE2001}.
\\
                                &  \FRBDMMWISMYMW \dmunits & 
Contribution of the Galactic ISM to $\DMFRB$ according to the YMW16 model \citep{YMW16}.
\\

$\DMMWHalo$                     & \FRBDMMWHalo \dmunits    &
Adopted contribution of the Galactic halo to $\DMFRB$.
\\
$\DMExgal$                      & \FRBDMEG \dmunits         &
Extragalactic contribution to $\DMFRB$, estimated by subtracting the adopted \DMMWISM{} (NE2001) and \DMMWHalo{}
\\ & & from \DMFRB.
\\
$E(B-V)_\mathrm{MW}$              & 0.0125 mag &
Galactic reddening along the FRB line-of-sight, from the reddening maps of \citet{SandF}.
\\
$P(U)$                          & 0.2                       &
Adopted PATH prior for the host being unseen.
\\
$P(U|x)$                        & $\sim 1$     &
PATH posterior for the host being unseen.

\end{tabular}
\end{table*}

\begin{figure}
    \centering
    \includegraphics[width=0.5\textwidth,trim={0 0.0in 0 0.0in},clip]{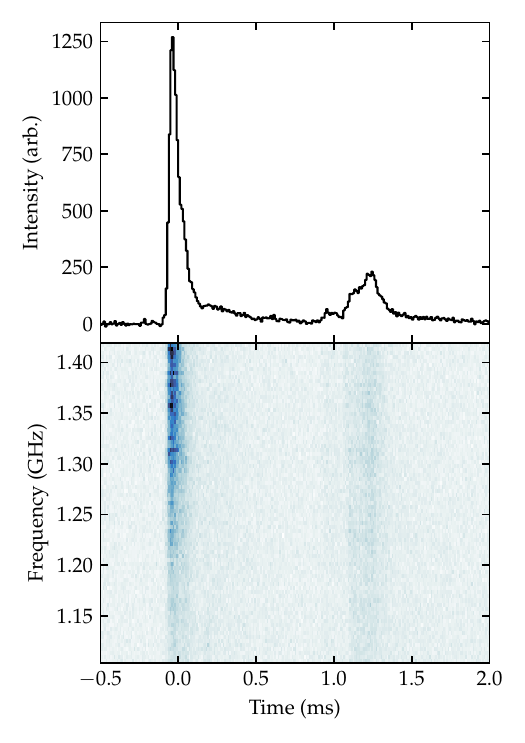}
    \caption{De-dispersed pulse profile (top) and dynamic spectrum (bottom) of \FRBname{}, at a temporal resolution of \SI{10}{\micro\second} (each sample summed from 3360 samples at \SI{3}{\nano\second} resolution) and spectral resolution of \SI{4}{\mega\hertz}.} % Courtesy of Danica Scott!
    \label{fig:pulse}
\end{figure}

\FRBname{} was detected during commensal observations with the Evolutionary Map of the Universe \citep[EMU;][]{EMU} project, observing at a central frequency of \FRBfreq{}~\si{\mega\hertz}. This triggered the CRAFT voltage-buffer downloads necessary for localisation. The burst arrived at UT 13:30:04.9, 2021-09-12 with a S/N of \FRBSN{}.
These data were reduced and processed using the standard CRAFT post-processing pipeline \citep{Day2020, Day2021}. 
Offline integration of the FRB signal over a window of six 1.182\,ms samples produced a fluence of $46.3\,\pm\,1.4$\,Jy\,ms, which implies a fluence of $69.9 \pm 2$\,Jy\,ms when accounting for the reduced sensitivity of the beam ($B=0.64$) at $0.475^{\circ}$ from centre, where the FRB was localised.

The voltage data were re-processed with the CRAFT Effortless Localisation and Enhanced Burst Inspection Pipeline \citep[CELEBI;][]{CELEBI}
to obtain high-time resolution data of the FRB pulse and localise 
\FRBname{} to $\alpha=$ \FRBRA{}, $\delta=$ \FRBDec{} (J2000), with a total 
uncertainty of \raerr, \decerr\ (RA, Dec; 68\%\ c.l.). This analysis found
$\DMFRB = \FRBDM\,\pm\,\FRBDMerr{}\,\dmunits$ using structure-maximisation techniques \citep{DMstruct}. Figure~\ref{fig:pulse} shows
a de-dispersed ``waterfall'' dynamic spectrum of the burst  and a collapsed profile at high time resolution.
A summary of the burst properties is given in \autoref{tab:quantities}.

\FRBname{} exhibits a complex multi-component pulse structure (apparent in \autoref{fig:pulse}), to be examined in detail in a forthcoming work (Bera et al, in preparation). For this paper, we take measurements from the initial, and brightest, pulse, on which the detection was triggered. We perform a scattering analysis using the methodology outlined by \citet{QiuBroadening}, as demonstrated by \citet{RyderFRB20220610A} and \citet{Sammons23}. We fitted a scattered Gaussian model to $250$ time samples surrounding the first component in the burst profile. We measure a narrow intrinsic width of $\widthintFRB=\SI{\FRBwidth}{\micro\second}$ and a scattering time of $\tauFRB=\SI{\FRBtau}{\micro\second}$ at a central frequency of $\FRBfreqtau$\,MHz. Following the method of \citet{Sammons23}, we divide the dynamic spectra into four sub-bands and independently fit the burst profile in each to measure the spectral evolution of $\tauFRB$. We find the evolution to be well described by a power law with a spectral index $\alpha=-4.7\pm0.4$, consistent with the evolution expected for scattering by Kolmogorov turbulence. The other components within the burst profile display much larger intrinsic widths and lower signal-to-noise ratios, and are therefore less constraining on the scattering time. 

\subsection{Optical and infrared follow-up}
\label{sec:optical}

Although \FRBname{} is located within the GAMA23 \citep{GAMA, GAMADR4} field, those observations are not of sufficient depth to be informative here. 

Upon localisation of the FRB signal, the coordinates were submitted for observation in $g$ and $I$ 
filters\footnote{\texttt{g\_HIGH} (filter ID ESO1115) and \texttt{I\_BESS} (ESO1077)}
on the FOcal Reducer and low-dispersion Spectrograph 2 \citep[FORS2;][]{FORS}, mounted on Unit Telescope 1 of the European Southern Observatory's Very Large Telescope (ESO VLT). With five dithered positions each, the total exposure time was \SI{2500}{\second} and \SI{450}{\second}, respectively. These observations were executed on 2021-10-04 UT under programme 105.204W.003 (PI Macquart). 

For the FORS2 images, de-biasing and flat-fielding were performed using the ESO Reflex \citep{ESOReflex} software package. The frames were coadded using Montage \citep{Montage} and \textsc{ccdproc} \citep{ccdproc}, and astrometric calibration was undertaken using the Astrometry.net \citep{Astrometry} code with indices generated from \textit{Gaia} DR3 \citep{GaiaDR3}. The coadded images were photometrically calibrated against DR2 of the DECam Local Volume Exploration survey \citep[DELVE;][]{DELVE} for $g$-band, and using the FORS2 Quality Control (QC1) archive for $I$ band, with Source Extractor \citep{SExtractor}, PSFEx \citep{PSFEx} and SEP \citep{SEP, SEP2} used for photometry. This procedure was chained together and tracked by the \textsc{craft-optical-followup} pipeline 
code\footnote{\url{https://github.com/Lachimax/craft-optical-followup/tree/marnoch+2023}}, 
making use of Astropy \citep{astropy1, astropy2}, NumPy \citep{numpy} and \textsc{astroquery} \citep{astroquery}. 
$E(B-V)=0.0125$, along the FRB line of sight, is taken from the \citet{SandF} reddening map (retrieved from the IRSA Dust 
Tool\footnote{\url{https://irsa.ipac.caltech.edu/applications/DUST/}} using \textsc{astroquery} \citep{astroquery}; 
we use this in conjunction with the \citet{F99} reddening law and assume $R_V=3.1$ to estimate the per-bandpass Galactic extinction $A_\lambda$ (given in \autoref{tab:limits}). We estimate the astrometric uncertainty of the imaging as the RMS of the offset of field stars from their counterparts in the \textit{Gaia} DR3 catalogue \citep{GaiaDR3}.

Optical observations are somewhat stymied by the presence of three 
stars %\footnote{With Gaia DR3 IDs 2329635654860257792, 2329647543329733760 and 2329647440250518528.} 
with \textit{Gaia} $G$-band magnitudes of 12.9, 10.9 and 9.5 \citep{GaiaDR3}, each $\sim1\arcmin$ from the FRB line-of-sight.
The scattered light from these stars produces a significant gradient at the FRB position, complicating photometry and host association. To mitigate this, we subtracted local background from each frame before coaddition. We first detect and mask objects using SEP \citep{SEP}; we then fit a third-degree two-dimensional polynomial to a 15-arcsecond square centred on the burst coordinates.
We selected a third-order polynomial as a compromise between over-fitting and background flatness. The background-subtracted portions of the images are shown in \autoref{fig:imaging}. Although successful in removing the gradient, shot noise from the stars makes our limits shallower than they could otherwise be; they are, nonetheless, relatively deep.

Our observing strategy was typical of that employed by CRAFT up to that point. 
However, for the first time, no host galaxy was identifiable by visual inspection. 

With the lack of a visible host, another set of FORS2 observations, with 18 dither positions totalling \SI{5940}{\second} (99 minutes) was requested using the $R$-band 
filter\footnote{\texttt{R\_SPECIAL}, filter ID ESO1076}, 
to reach a greater depth. The pointing was specified to position the two brightest of the nearby stars outside of the field of view. This OB was executed on 2021-10-09 UT. These images were processed using the same procedure as those in $g$ and $I$-band, with photometric calibration sourced again from the QC1 archive. Although the scattered light from these stars remains substantial, a greater 5-$\sigma$ depth of $\Rmaglim$ AB mag was achieved, using an aperture radius of twice the PSF FWHM of $\Rfwhm$\,arcsec. Nonetheless, no object became apparent.

Typically, galaxies are brighter in the near-infrared than the optical, with cosmological redshift pushing even more light outside of the visible range. With the possibility of a high-redshift host in mind, a further observation was requested with the High Acuity Wide-field $K$-band Imager \citep[HAWK-I;][]{HAWK-I}, a near-infrared instrument mounted on Unit Telescope 4 of the VLT (PI Shannon, programme 108.21ZF.005). HAWK-I was used in AutoJitter mode in conjunction with the GRAAL ground-layer adaptive optics system, specifying 
15 $\times$ 10-second integrations per offset, with 16 pseudo-random offsets within a 
16\,arcsec jitter box, totalling 
\SI{2400}{\second} of integration time. This was first executed on UT 2021-12-07, but the seeing constraints were exceeded and the observation was repeated on 2021-12-13. As the first observation set was not strongly impacted by the breach of constraint, we combined both observations for double the total integration time.
We used a similar procedure to the FORS2 imaging to process the HAWK-I imaging, but with ESO Reflex taking on the coaddition. Photometric calibration was performed with reference to the 2MASS Point Source Catalog \citep{2MASS}, and again \textit{Gaia} DR3 was used for astrometric calibration. Once more, no host was noted to a depth of \Kmaglim{} AB mag.

These observations are summarised in \autoref{tab:limits}. The combination of $R$ and $K_s$ imaging, used to guide spectroscopic observations from X-Shooter, now forms the basis of the CRAFT follow-up strategy in the form of the Fast and Unbiased FRB host galaxY survey (FURBY), an ongoing Large Programme at the VLT.  However, we do not have spectroscopic information for this field.

\section{Analysis}
\label{sec:analysis}

\subsection{Attempting host association}
\label{sec:path}

\begin{table*}
    \centering
    \caption{Photometric limits and other properties for observations targeting the host of \FRBname.  We also give, in $m_\mathrm{faintest}$, the magnitude of the faintest object considered by PATH which has also been detected in at least one other band; generally these are fainter than the stated statistical limit, but with less than 3-$\sigma$ significance. The bracketed letter next to the magnitude identifies the PATH candidate, as listed in \autoref{tab:path}. $A_\lambda$ is the estimated Galactic extinction for that band along the line-of-sight (see \autoref{sec:optical}).
    $\sigma_\mathrm{astm}$ gives the RMS of the offset of field stars in the imaging from their \textit{Gaia} \citep{GaiaDR3} counterparts.}
    \begin{tabular}{llccccccccc}
        Instrument  & Band                  & Observation           & Integration          & PSF       & $\sigma_\mathrm{astm}$& $5\sigma$     & $m_\mathrm{faintest}$ & $A_\lambda$   \\
                    &                       & initiated             & time              & FWHM      &                       & lower limit   &                       &               \\
                    &                       & (UT)                  & (\si{\second})    & (arcsec) & (arcsec)             & (AB mag)      & (AB mag)              & (AB mag)
        \\\hline
        \\
        VLT/FORS2   & $g_\mathrm{high}$     & 2021-10-04T03:34:40   & $2500$            & $\gfwhm$  & 0.21                  & $\gmaglim$    & 27.2 (A)              & $\gext$
        \\
        VLT/FORS2   & $R_\mathrm{special}$  & 2021-10-09T02:19:36   & $5840$            & $\Rfwhm$  & 0.097                 & $\Rmaglim$    & 26.9 (A)        & $\Rext$
        \\
        VLT/FORS2   & $I_\mathrm{Bess}$     & 2021-10-04T04:20:26   & $450$             & $\Ifwhm$  & 0.24                  & $\Imaglim$    & 25.5 (E)              & $\Iext$
        \\
        VLT/HAWK-I  & $K_\mathrm{s}$        & 2021-12-07T01:42:23   & $4800$            & $\Kfwhm$  & 0.11                  & $\Kmaglim$    & 22.9 (H)      & $\Kext$
        \\          &                       & 2021-12-13T01:17:38   &         
        
    \end{tabular}
    \label{tab:limits}
\end{table*}

% Faintest multi-band PATH object in this band
% g: 27.366029396976323
% R: 26.976353887868733
% I: 25.513931794610993
% K: 21.080968183686213

        % VLT/FORS2   & $g_\mathrm{high}$     & 2021-10-04T03:34:40   & $2500$            & $\gfwhm$  & 28.1      & 26.9      & $\gmaglim$    & 27.3 (A)              & $\gext$
        % \\
        % VLT/FORS2   & $R_\mathrm{special}$  & 2021-10-09T02:19:36   & $5840$            & $\Rfwhm$  & 28.5      & 27.3      & $\Rmaglim$    & 27.0 (A)              & $\Rext$
        % \\
        % VLT/FORS2   & $I_\mathrm{Bess}$     & 2021-10-04T04:20:26   & $450$             & $\Ifwhm$  & 26.0      & 24.8      & $\Imaglim$    & 25.5 (E)              & $\Iext$
        % \\
        % VLT/HAWK-I  & $K_\mathrm{s}$        & 2021-12-07T01:42:23   & $4800$            & $\Kfwhm$  & 24.9      & 23.7      & $\Kmaglim$    & 22.3 (I)              & $\Kext$

\begin{figure}
    \centering
    \includegraphics[width=0.48\textwidth,trim={0 0.0in 0 0.0in},clip]{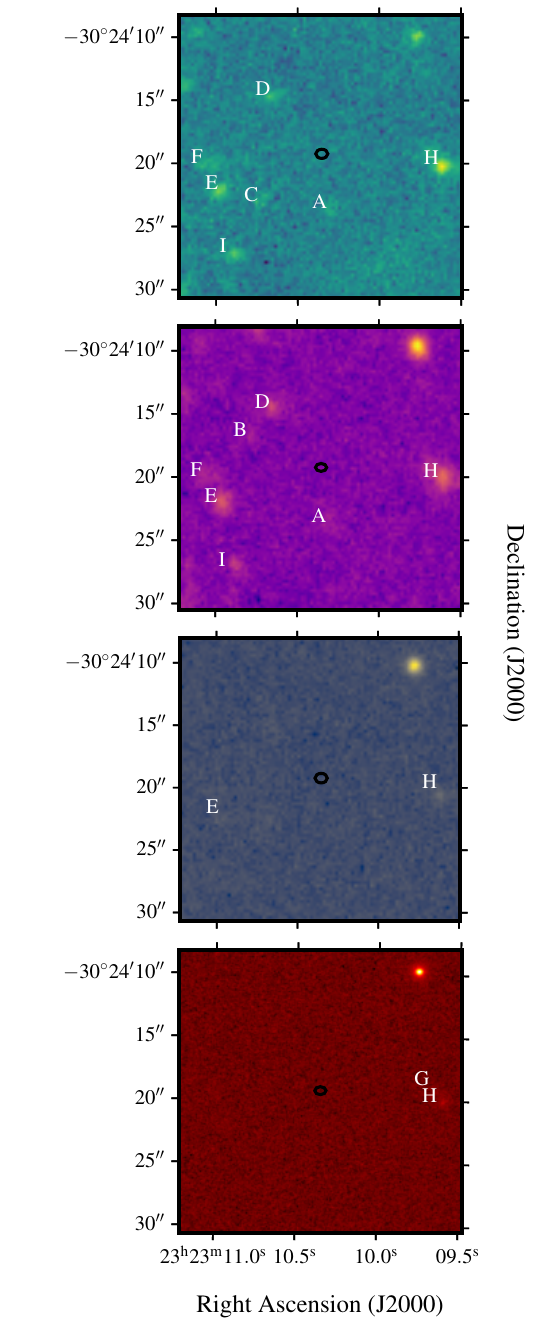}
    \vspace{-0.05in}
    \caption{Background-subtracted VLT imaging at the position of \FRBname{}. From top to bottom: FORS2/$g$-band, FORS2/$R$-band, FORS2/$I$-band, and HAWK-I/$\Ks$-band.
    Letters denote PATH candidates detected in each image (see \autoref{tab:path}); a missing letter indicates that the given source was not detected in that band.
    The black ellipse in each image outlines the localisation region of the FRB, combined in quadrature with the image astrometric uncertainty.
    }
    \label{fig:imaging}    
\end{figure}

As standard practice for CRAFT FRB detections, we attempt to associate the burst to a host galaxy using the Probabilistic Association of Transients to their Hosts \citep[PATH;][]{PATH} code. The PATH methodology uses Bayesian inference to assign a probability to each nearby object in the field having hosted the transient, taking the positions, magnitudes, and angular sizes of those objects as inputs, as well as the FRB localisation region. 
The \textsc{frb} \citep{FRBrepo}
repository\footnote{\url{https://github.com/FRBs/FRB}}
offers a convenient wrapper for PATH which derives the relevant measurements from an imaging FITS file before passing the resulting catalogue to PATH.
This uses image segmentation, via \textsc{photutils}, to perform photometry, which tends to over-attribute flux to some objects in the presence of a background gradient. For this reason we apply PATH after the scattered light from the nearby bright stars has been subtracted, and the images trimmed to the subtracted region. 

Typically, PATH is applied only to one deep image of the field, preferably in $r$-band; for the sake of completeness, we run it on all four of our images. We note that our $R$-band imaging satisfies all of the recommended criteria of \citet{PATH}, namely a 5-$\sigma$ depth greater than 25.5 mag and seeing less than 1 arcsec.
We also incorporate the astrometric uncertainty of the image (given in \autoref{tab:limits}) by adding it in quadrature with the FRB positional uncertainty.

We adopt the \scarequotes{exponential} prescription (with the maximum offset set to the default of 6 half-light radii) for calculating the host radial offset prior $p(\omega|O_i)$ and the \scarequotes{inverse} prescription for the candidate priors $P(O_i)$, as described by \cite{PATH}. For the prior for the true host being unseen we adopt $P(U) = 0.2$, given the large DM (see Appendix \ref{app:p_u}). 
We summarise the 9 host candidates identified by PATH within the radial cutoff of 11 arcsec, across all bands, in \autoref{tab:path}, with the full results in each band given in Appendix \ref{app:path}. The candidates are also labelled in the imaging in \autoref{fig:imaging}. The nearest source, Candidate A, is 4.4\,arcsec from the FRB sightline; it is detected in $g$ and $R$ bands, in both of which it is assigned the greatest $P(O_i|x)$, the posterior probability that it is the FRB host ($g$: $9.2\times10^{-39}$; $R$: $3.6\times10^{-63}$).
The resulting posterior of the true host being unseen in $R$-band, $P(U|x)$, differs from unity by only $4\times10^{-63}$.

To examine the influence of our priors, we vary $P(U)$ as low as 0.01, in which case the $R$-band $P(U|x)$ is lower, but at minimum is only less than unity by $10^{-61}$. For further discussion of this prior, see Appendix \ref{app:p_u}. 
We also investigated the effect of adopting different priors $p(\omega|O_i)$, for the host radial offset. We trialled the \scarequotes{core} and \scarequotes{uniform} prescriptions (with maximum offset kept at 6 half-light radii and holding the other priors as adopted) and found negligible effect, with $P(U|x)$ differing from unity by only $2\times10^{-60}$ and $8\times10^{-60}$ respectively.
Thus we conclude that the true host is not present in the image.

\begin{table}
\centering
\caption{PATH \citep{PATH} host galaxy candidates for \FRBname{}, derived from VLT imaging with the priors given in \autoref{sec:path} and radial cutoff 
\PATHradius; as labelled in \autoref{fig:imaging}. They are organised by $R_\perp$, the angular distance from the FRB line-of-sight. %`Max $P(O_i|x)$' is the maximum posterior probability for that object, and `Max band' is the band in which that maximum is reached. 
All candidates have posteriors $P(O_i|x) < 10^{-38}$ in each band, which we consider negligible. 
Full PATH outputs, including photometry, are given in Appendix \ref{app:path}}.

\label{tab:path}
\begin{tabular}{cccccccc}
ID  & $\alpha$  & $\delta$  & $R_\perp$       \\              %& Max           & Max  \\% & $P_R(O_i|x)$ & $P_g(O_i|x)$ & $P_I(O_i|x)$ & $P_K(O_i|x)$ \\
    & (J2000)   & (J2000)   & (arcsec)         \\            % & $P(O_i|x)$    & band    \\%&  &  &  &  \\
\hline \\
A & $23^{\mathrm{h}}23^{\mathrm{m}}10\rlap{.}^\mathrm{s}30$ & $-30^\circ24{}^\prime23\rlap{.}^{\prime\prime}6$ & 4.4 % & $9 \times 10^{-39} \; \mathrm{}$ & $g$ 
\\
B & $23^{\mathrm{h}}23^{\mathrm{m}}10\rlap{.}^\mathrm{s}77$ & $-30^\circ24{}^\prime16\rlap{.}^{\prime\prime}8$ & 6.0 %& $0 \; \mathrm{}$ & $R$
\\
C & $23^{\mathrm{h}}23^{\mathrm{m}}10\rlap{.}^\mathrm{s}71$ & $-30^\circ24{}^\prime23\rlap{.}^{\prime\prime}0$ & 6.0 %& $2 \times 10^{-78} \; \mathrm{}$ & $g$
\\
D & $23^{\mathrm{h}}23^{\mathrm{m}}10\rlap{.}^\mathrm{s}64$ & $-30^\circ24{}^\prime14\rlap{.}^{\prime\prime}6$ & 6.0 %& $2 \times 10^{-74} \; \mathrm{}$ & $g$
\\
E & $23^{\mathrm{h}}23^{\mathrm{m}}10\rlap{.}^\mathrm{s}95$ & $-30^\circ24{}^\prime22\rlap{.}^{\prime\prime}1$ & 8.2 %& $1 \times 10^{-184} \; \mathrm{}$ & $I$
\\
F & $23^{\mathrm{h}}23^{\mathrm{m}}11\rlap{.}^\mathrm{s}04$ & $-30^\circ24{}^\prime20\rlap{.}^{\prime\prime}0$ & 8.9 %& $2 \times 10^{-173} \; \mathrm{}$ & $g$
\\
G & $23^{\mathrm{h}}23^{\mathrm{m}}09\rlap{.}^\mathrm{s}66$ & $-30^\circ24{}^\prime19\rlap{.}^{\prime\prime}4$ & 9.0 %& $0 \; \mathrm{}$ & $K$
\\
H & $23^{\mathrm{h}}23^{\mathrm{m}}09\rlap{.}^\mathrm{s}61$ & $-30^\circ24{}^\prime20\rlap{.}^{\prime\prime}0$ & 9.6 %& $4 \times 10^{-137} \; \mathrm{}$ & $g$
\\
I & $23^{\mathrm{h}}23^{\mathrm{m}}10\rlap{.}^\mathrm{s}86$ & $-30^\circ24{}^\prime27\rlap{.}^{\prime\prime}1$ & 10.2 %& $0 \; \mathrm{}$ & $R$
\end{tabular}
\end{table}

\subsection{Magnitude limits}

Our magnitude limits, given in \autoref{tab:limits}, are derived from the background-subtracted versions of the images. It should be noted that, even after this subtraction, Poisson noise from the bright stars will reduce the depth of the images at the FRB position; nonetheless, we attain sensitive photometric limits.
These are calculated using RMS error maps generated by SEP \citep{SEP}; we sum the square of the RMS assigned to the pixels within the circular 
aperture\footnote{A similar process is used internally by SEP and Source Extractor to derive their photometric uncertainties.}
and take the square root of the sum; this is the $1\sigma$ flux limit. This is multiplied by the appropriate factor and converted into a magnitude.

In each image, we use a circular aperture with a radius of twice the delivered PSF FWHM at the centre of the FRB uncertainty ellipse.
We find a 5-$\sigma$ AB magnitude
limit of $R > \Rmaglim$\,mag and $K > \Kmaglim$\,mag. The host of \FRBname{} must be considerably fainter than any FRB host known; the faintest to date, the host of \FRB{20121102A}, has an $r$-band AB magnitude, corrected for Galactic extinction, 
of 23.73\,mag \citep{GordonProspector, Bassa2017}, almost 3 mag brighter than our limit.

\subsection{Redshift estimation}
\label{sec:p_z_dm}

\begin{figure*}
    \centering
    \includegraphics[width=\textwidth]{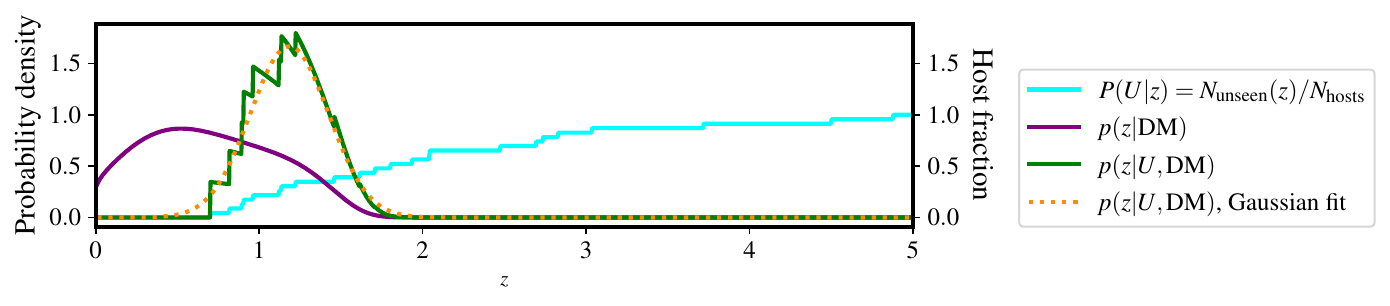}
    \caption{
    The cyan line traces the fraction of hosts in our SED sample that would be fainter than the magnitude limit of our $R$-band imaging at the given redshift, $N_\mathrm{unseen}(z) / N_\mathrm{hosts}$; we use this as a coarse probability that a new FRB host at redshift $z$ would be unseen, or $P(U|z)$.
    The purple line is the best-fitting $p(z|\DM{})$ probability density function for \FRBname{}.
    The green line is the product of these, giving a posterior distribution for the redshift of the host of \FRBname{}, assuming it is like the hosts in our sample, while the dashed orange line is a Gaussian fit to the former.
    }
    \label{fig:pdfs}
\end{figure*}

We derive a probability density function $\pzdm$
for the cosmological redshift of the source \FRBname{} using the {\sc zDM} analysis of \citet{James2022A}. We assume the best-fitting cosmology from \citet{Planck2018} (in particular, $H_0 = 67.4$\,km\,s$^{-1}$\,Mpc$^{-1}$), 
and FRB population parameters from \citet{JamesH0}, updated for a new FRB maximum energy from \citet{RyderFRB20220610A}. 
We calculate an extragalactic DM contribution, $\DMExgal$, 
of $\FRBDMEG$\,\dmunits, after subtracting an assumed Milky Way halo contribution of 50\,\dmunits, and the line-of-sight estimate of \FRBDMMWISMNE\,\dmunits\ from the interstellar medium (ISM) as predicted by the NE2001 
model\footnote{The YMW16 \citep{YMW16} prediction for this line-of-sight is only $\FRBDMMWISMYMW{}$ \dmunits; fortunately, the difference between the two models constitutes only a 1\% effect on \DMFRB{}.}
\citep{NE2001}. 
Rather than average over the FRB width $w_\mathrm{FRB}$ and signal-to-noise (S/N) distributions as in the usual {\sc zDM} analysis, we use the exact FRB parameters for \FRBname{}: detection $S/N=\FRBSN{}$, and $w=\FRBwidthtotal{}$\,ms (including scattering and intrinsic width). Thus we produce $p(z|\DM{})$ for this specific FRB, as was done for \citet{Lee-WaddellFRB20171020A}, rather than the overall population.

The best-fitting curve is shown in \autoref{fig:pdfs}.
The burst's high S/N of \FRBSN{} skews the distribution toward lower redshifts but does not exclude high redshift values up to $\sim1.8$; this FRB could be an outlier in terms of luminosity. In the absence of deep imaging, we would conclude that the FRB was most likely emitted at a redshift of $\zhost\sim0.5$, high for the CRAFT sample but similar to those of FRBs 20181112A and 20190711A; and that the large DM excess is probably attributable to the host ISM or source environment. However, the lack of any host galaxy in such deep imaging severely challenges this interpretation. 

While we have used the best-fitting model here, variation in FRB population parameters within their known uncertainties will produce different results. Confirmation of a large redshift for this host will rule out that part of the parameter space predicting a low $z$, and constrain FRB populations further.
Such a confirmation could further increase the maximum FRB energy $E_{\rm max}$, which is only a few orders of magnitude below theoretical limits \citep{LuKumar2019}; improve the chances for constraining high-redshift phenomena such as helium reionisation with future instruments \citep[see e.g.\ ][]{Caleb2019a}; and point towards FRBs having very high host DMs
being the exception--perhaps due to being a separate population from young magnetars--rather than the rule.

\subsection{Modelling of other hosts}
\label{sec:sed}

\begin{figure*}
    % \begin{minipage}{\textwidth}
    \centering
    \includegraphics[width=\textwidth]{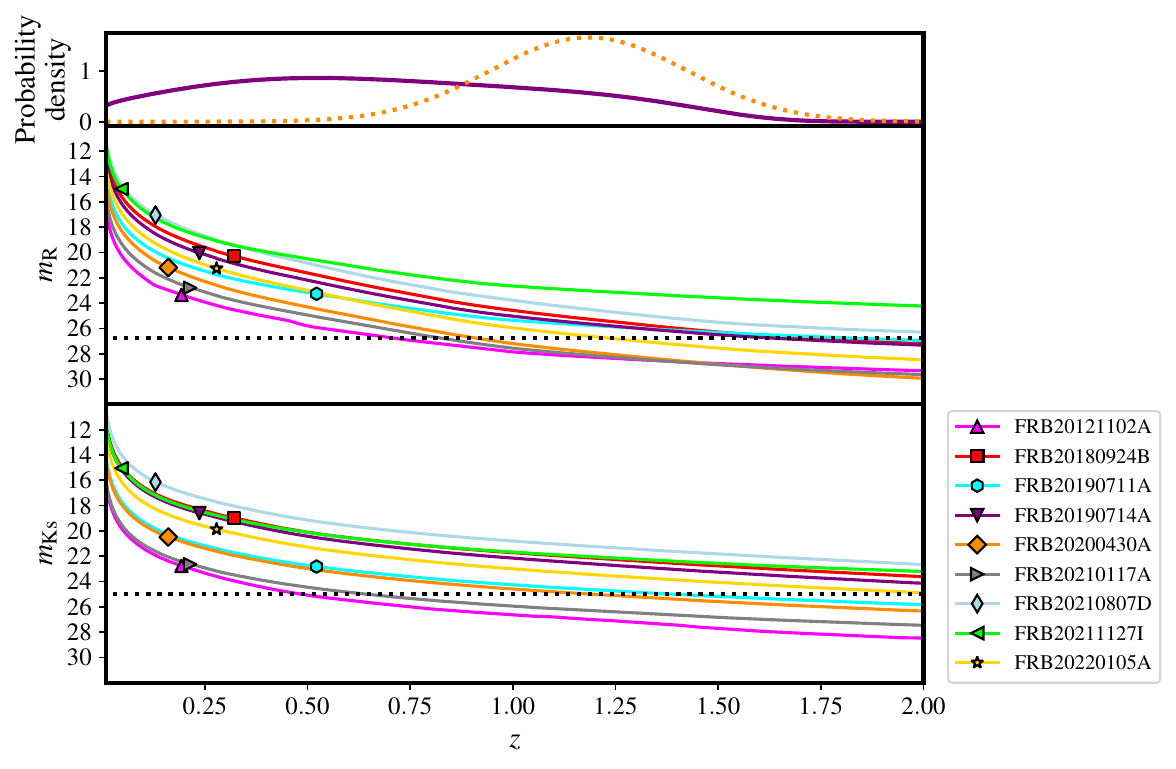}
    \caption{
    Redshift-magnitude diagrams (top panel: $R$-band, bottom panel: $\Ks$-band) diagrams for 9 FRB hosts, selected to span the luminosity distribution of our sample of SED models (described in \autoref{sec:sed}), including the most and least luminous. The full sample is visualised in \autoref{fig:zm_all}. The solid lines trace the integrated apparent magnitude that would be observed for an FRB host if it were placed at the given redshift, with a marker placed at its true measured redshift. The black dotted line marks the 5-$\sigma$ limit given in \autoref{tab:limits}. 
    The top panel provides $\pzdm{}$ (purple line) and the Gaussian fit to $p(z|U,\DM{})$ (orange dotted line), as also shown in \autoref{fig:pdfs}}.
    \label{fig:zm}
    % \end{minipage}
\end{figure*}

\begin{figure}
    % \begin{minipage}{0.4\textwidth}
    \centering
    \includegraphics[width=0.5\textwidth]{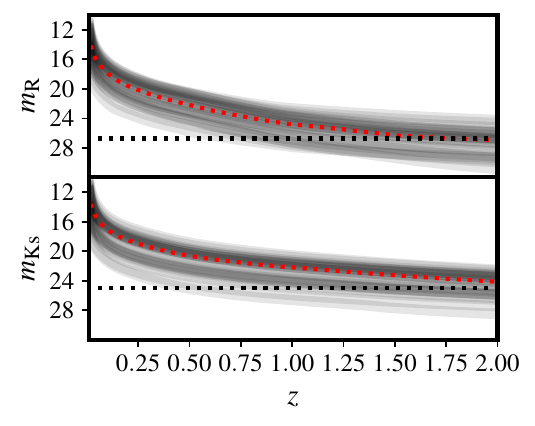}
    \caption{
    Redshift-magnitude diagrams (top panel: $R$-band, bottom panel: $\Ks$-band) for all 23 FRB hosts in the SED model sample, with lines left unlabelled but semi-transparent.
    The red dotted line gives the median magnitude at each redshift.
    The black dotted line marks the 5-$\sigma$ limit given in \autoref{tab:limits}.
    }
    \label{fig:zm_all}
    % \end{minipage}
\end{figure}

\begin{figure}
    % \begin{minipage}{0.4\textwidth}
    \centering
    \includegraphics[width=0.5\textwidth]{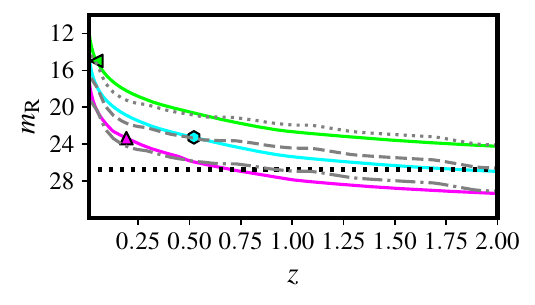}
    \caption{
    Redshift-magnitude ($z$-$m_\mathrm{R}$) diagram for three FRB hosts (coloured lines and symbols - FRB legend is given in \autoref{fig:zm_all}). 
    The grey lines show the expected brightness for $L^*$, 0.1 $L^*$, and 0.01 $L^*$ galaxies at these redshifts. 
    The black dotted line marks the 5-$\sigma$ $R$-band limit given in \autoref{tab:limits}.
    }
    \label{fig:rmag}
    % \end{minipage}
\end{figure}

In order to compare this FRB with others in which the host galaxy is known, we have  simulated their detectability when placed at a range of redshifts. We employ a set of SED models derived from joint fitting of photometry and spectroscopy using the stellar population synthesis code \textsc{prospector} \citep{Prospector, ProspectorSource}, as described by \cite{GordonProspector}. All 23 hosts modelled are considered to be highly secure ($P(O_i|x) > 0.9$) identifications; included are 17 identified by CRAFT and six by other groups, constituting the majority of reported FRB hosts at the time of submission. We use these models to estimate the brightness of each host when placed at a range of redshifts.
At each $z$, we shift the wavelength coordinates and attenuate the observed flux as appropriate 
(following equations provided by \citealp{HoggCosmology}), and use the VLT-FORS2 $R_\mathrm{special}$ bandpass to derive the hypothetical flux and AB magnitude for each galaxy as a function of (displaced) redshift. The magnitude-redshift diagrams for nine of these models, selected to span the luminosity distribution of our SED model sample, are shown in \autoref{fig:zm}. We also plot the entire sample in \autoref{fig:zm_all}.
It should be made clear that we are not attempting to rewind these galaxies to the states they were in at the look-back time of the redshift in which we place them; rather, we answer the somewhat counterfactual question of how they would appear if transported, as they are presently observed and without changes, to the redshift (that is, both the cosmological distance and look-back time) in question. It is also worth noting that this approach does not account for the fact that faint structures in the hosts contribute to the flux at low redshift, but may not pass the detection threshold at higher redshifts; thus the actual measured magnitude from a galaxy placed at a higher redshift may be fainter than expressed here, and the detectability of our hosts at higher redshifts may be slightly exaggerated. However, we believe it suffices as an approximation. We recognise that, although representing the majority of known FRB hosts, this sample may not be unbiased. In particular, the selection cuts made by \citet{GordonProspector} favour brighter galaxies, and the adopted PATH priors therein may also; see \citet{GordonProspector} for a discussion. However, we assert that these biases are likely to be minor.
We use the detectability of these model hosts to frame our discussion of the host scenarios below.

For context, we have plotted again a subset of these in \autoref{fig:rmag}, with overplotted estimates for the characteristic 
galaxy luminosity $L^*$ based on a series of
wide-field galaxy surveys at these redshifts \citep{Brown2001, Wolf2003, Willmer2006, Reddy2009, Finkelstein2015, Heintz2020}, while recognising that $L^*$ is a characteristic of the galaxy luminosity function, which itself evolves with redshift. Hence these $L^*$ curves do not follow the exact same evolution with $z$ that our hosts 
do. 
The other FRB hosts span a range of luminosities relative to
$L^*$, with the majority
sub-$L^*$ but brighter than $0.1L^*$.

\section{Host scenarios \& Discussion}
\label{sec:scenarios}

\subsection{Case 1: Visible host at large offset}

Although there is no stand-out host candidate in our imaging, there are other objects nearby in the field-of-view (see \autoref{fig:imaging} and \autoref{tab:path}). We consider the possibility that one of these is the host of \FRBname{}.
The nearest to the line-of-sight, Candidate A, is separated by 4.4\,arcsec to the south-west from the FRB coordinates. 
Unfortunately, we do not have spectroscopy for any of the objects considered, and photometric redshifts are impractical at the S/N and sparseness of photometric data-points of these objects.
The maximum projected distance (at the cosmological angular size turnover point, $z\sim1.5$) for this object is 39\,kpc. This is not an unreasonable offset for an object in a massive galaxy, particularly if it happens to lie in a globular cluster \citep[e.g.][]{GlobularCluster}; however, given the faintness of this object ($R=27$), it is unlikely to extend to such distances. A similar argument applies to each nearby object. %Given their small angular sizes, it is not clear whether they are even resolved, and, lacking spectroscopy, we cannot rule out whether some are faint stars within the Milky Way.
We thus conclude that no detected object is likely to be the host galaxy, consistent with PATH posteriors.

\subsection{Case 2: Unseen host at modest redshift (< 0.7)}
\label{sec:low-z}

The peak likelihood of $\pzdm$ (the purple curve in \autoref{fig:pdfs}) is at $z=0.52$, and the current average FRB host sits at $z\sim0.2$. However, if the host of \FRBname{} was at either of these redshifts, it would be easily detectable in our $R$-band imaging unless it was significantly less luminous than any known FRB host.
As illustrated in \autoref{fig:zm}, the host galaxy of the first known repeating FRB, \FRB{20121102A} (henceforth R1), would only become undetectable in this imaging if placed above redshift 0.7. Hence, if our host is similar to the R1 host, it would still place the \FRBname{} host at a higher redshift than any other known except for \FRB{20220610A}. 
If it is \textit{less} luminous than the R1 host, it would become the dimmest FRB host known.
Some dwarf FRB host galaxies have
%even
been shown to have unusually high 
$\DMHost$ contributions \citep{SimhaExcess}; 
it is interesting that three of them are the dwarf hosts mentioned previously, that is: FRBs 20121102A ($\zhost=0.193$, $\DMFRB\sim555\dmunits$; \citealp{HostR1}); 20210117A ($\zhost=0.2145$, $\DMFRB{}=728.95\dmunits$; \citealp{BhandariFRB20210117A}); and 20190520B ($\zhost=0.241$, $\DMFRB\sim1204\dmunits$; \citealp{R1-twin}). 
Although it is quite conceivable that our missing FRB host is a dwarf at a relatively modest redshift, a higher redshift is necessary if the host is within the parameters of the known host population. With reference to the SED models described in \autoref{sec:sed}, only five out of 23 FRB hosts are lost at $z<1$ in $R$-band. 
A lower redshift, as the event's brightness suggests, lies in tension with the lack of an observed host; 
if $\zhost<0.7$, it must be less luminous than any known FRB host. 

\subsection{Case 3: Unseen host at high redshift (>0.7)}
\label{sec:high-z}

\pzdm{} extends to redshifts significantly greater than 1.  In addition to this, using the sample of SED models described in \autoref{sec:sed}, we produce a coarse probability that a new FRB host would be unseen in our $R$-band imaging, as a function of its redshift, by simply taking the fraction of the sample at each given redshift fainter than our imaging limits, i.e. $P(U|z) = N_\mathrm{unseen}/N_\mathrm{hosts}$. When multiplied with the $\pzdm$ PDF, and normalised so that the total probability is 1, the green line in \autoref{fig:pdfs} is produced. We interpret this as a posterior for the host's redshift, given the properties of the pulse as well as the fact that it is unseen: 
$p(z|U,\DM{}) = P(U|z)\pzdm{}/P(U)$.
The normalisation factor here, given as $P(U)$, also has applications as a prior for use with PATH; see Appendix \ref{app:p_u}. To produce a smoothed variant, we fit a Gaussian distribution to the curve prior to normalisation (in $R$), and then normalise this PDF. This is the dashed orange line in \autoref{fig:pdfs}, with $\mu_\mathrm{U}=1.18$ and $\sigma_\mathrm{U}=0.24$. 
This allows a less ambiguous identification of the most likely redshift, and the low-$z$ wing allows for FRB hosts less luminous than in our currently-known sample. With this approach, considering all properties of the burst and detector (encoded in $\pzdm$) and the distribution of FRB hosts to date (collapsed into $P(U|z)$), it appears more probable that \FRBname{} is located at relatively high redshift.

The FRB signal itself exhibits low levels of scatter broadening, with 
$\tauFRB = \FRBtau\,\si{\micro\second}$.
This small $\tauFRB$ provides evidence for a small \DMHost{}, and hence a large source redshift.
Adopting the framework of \citet{Cordes2022}, with conservative estimates of the parameters defined 
there\footnote{$A_\tau \widetilde{F} G = \SI{0.1}{\parsec^{-2/3}\ \km^{-1/3}}$, 
with larger values equating to smaller \DMHost{}},
we find that \DMHost{} is likely $< 300\ \dmunits$. This leaves $> 850\ \dmunits$ for \DMCosmic{}, and places \FRBname{} at $\zhost\gtrsim0.9$ for an average sightline. 

If the event did occur at $z>1$, it could provide an additional high-redshift anchor 
to the Macquart relation and support the previously 
unanticipated
FRB energetics implied by \FRB{20220610A}. 
Placing the FRB at $z\approx1$ requires a rest-frame burst energy of $9.7 \times 10^{41}~\si{\erg}$, 
assuming a standard bandwidth of 1\,GHz in the burst rest frame and given the intrinsic fluence of 
$\FRBfluence{}\pm2$\,Jy\,ms. If we normalise this to the emission frequency of 1272.5\,GHz using a $k$-correction corresponding to a frequency dependence of 
$F_\nu \propto \nu^{-1.5}$, we find $E_{1272.5\,{\rm MHz}}=2.7 \times 10^{42}~\si{\erg}$.
This energy is in excess of the value of $E_{1271.5 {\rm MHz}} = 2.0 \times 10^{42}$\,erg found for \FRB{20220610A} \citep{RyderFRB20220610A}, making it one of the most energetic FRBs yet detected, and challenging existing estimates of the maximum burst energy and, hence, progenitor models \citep{LuKumar2019}. In turn, it would prove that ASKAP, and in particular its next-generation FRB-hunting system CRACO, will be able to detect FRBs up to $z \sim 3$ (i.e., over the peak of star-forming activity in the Universe), testing their hypothesised association with high-mass star formation against the alternative of collisions of stellar remnants \citep{MargalitMetzger2018,Totani2013}. It would also indicate that more sensitive systems such as FAST might be able to probe far enough back in time to detect the signature of helium reionisation
\citep{Caleb2019a}.

\subsection{Case 4: Orphan progenitor}

We consider the possibility of the FRB progenitor being truly hostless, having been ejected from its original galaxy. Such an outcome can be plausibly triggered by, for example, the gravitational interaction of stars in dense, multiple stellar systems like globular clusters \citep{CabreraRodriguez2023} or the interaction of a multi-star system with a massive or supermassive black hole \citep{Evans2022}. The latter case has been shown to occur for massive stars cast out of the Milky Way nucleus at high velocity \citep{Evans2022, Koposov2020}. Recently, a linear star-formation feature, believed to be non-galactic and to have been induced by the passage of an ejected supermassive black hole was identified \citep{RunawaySMBH}, although this interpretation has been challenged \citep{AlmeidaOrIsIt}. The existence, in galaxy clusters, of intergalactic planetary nebulae \citep{IntergalacticPNe}, supernovae \citep{IntergalacticSNe} and red-giant branch stars \citep{IntergalacticRGB} also suggests the possibility of intracluster FRB progenitors, although since we detect no evidence of a galaxy cluster in this field, this is unlikely to be the case for this FRB.

A hostless FRB progenitor would have $\DMHost\sim0$, 
and thus very likely be at $z>1$ for this FRB; the best-fit estimate from $\tavg{\DMCosmic}$ 
is $z\sim1.3$ (as estimated with tools provided in the \textsc{frb} code). Assigning this redshift to the nearest object (host candidate A), the separation of the burst position from the object centre \citep[galactic nuclei generally being where hypervelocity ejections begin;][]{Hills1988, Evans2022} projects to 38\,kpc. An ejected object travelling at \SI{1000}{\kilo\metre~\second^{-1}} \citep[a typical speed for a hypervelocity star ejected from the Milky Way via black hole interaction;][]{Evans2022, Koposov2020, Sherwin2008} would require 37~Myr to traverse this distance. While a neutron star would certainly survive this journey, a typical non-binary magnetar, with a lifespan of order $10^5$ years \citep{Magnetars}, would long have spun down and ceased activity. 
Although it appears that some FRBs originate from old stellar populations, as suggested by the M81 globular cluster repeater \citep{GlobularCluster} 
and the quiescent host of \FRB{20220509G} \citep{SharmaFRBClusters}, evidence is building that the majority trace star formation without a large delay \citep{JamesH0, GordonProspector}, thus favouring young objects as progenitors. 

As we have already discussed, a low-luminosity galaxy at this redshift could escape detection, thus an unseen galaxy at a smaller projected distance could have been the origin of the hypothetical rogue progenitor. A progenitor still bound to an unseen host galaxy is the more likely scenario.

\subsection{Case 5: Intrinsically hostless progenitor mechanism}

There are hypothesised FRB progenitor mechanisms that predict FRBs completely devoid of progenitor objects, although none are favoured within the wider community; for example, the decay of cosmic string cusps \citep{Brandenberger2017}. The thus-far consistent association of well-localised FRBs with host galaxies makes such models unlikely for the population at large, but does not rule them out as the cause of a small subset of bursts. As we can neither exclude nor provide further evidence for this scenario, we mention it here only for the sake of completeness.

%%%%%

\subsection{Continuing the hunt for the host}
\label{sec:future}

Either the host of \FRBname{} is dimmer than any known FRB host galaxy (as shown in \autoref{sec:low-z}), or it has one of the highest redshifts (potentially \textit{the} highest; see \autoref{sec:high-z}). Confirming which is of great interest, and either would provide an interesting new case study with value to investigations of either the FRB host population or cosmological parameters and the properties of the FRB population itself. To do so requires further observations. According to the FORS Exposure Time 
Calculator\footnote{\url{https://www.eso.org/observing/etc/}}, 
a starburst host \citep[Kinney Starburst 2 template;][]{KinneyTemplates} placed at $z=1.2$ with near-ideal conditions of 0.8\,arcsec delivered image quality, airmass 1.0, and a lunar illumination of 0.1, as well as assuming an unresolved source with brightness of $R=27.3$\,mag (which corresponds to the 3-$\sigma$ limit of our $R$-band imaging),
requires only 
1~hour of exposure to reach $S/N=5$ in $R$-band; however, this time does not account for noise provided by the nearby stars, and we must consider the possibility of a much fainter host. To reach $S/N=5$ for an $R=28.5$ (corresponding to our 1-$\sigma$ limit) galaxy under the same conditions requires nearly 9 hours of exposure time. With more realistic conditions similar to our actual observations, these times go to $\sim$2.5~hours for $R=27.3$\,mag and nearly 22~hours for $R=28.5$\,mag.

HAWK-I is similarly prohibitive, with 10,000~s of total integration required to reach $K=25.5$ (our 3-$\sigma$ limit) and 92,000~s for $K=26.7$\,mag (1-$\sigma$), with delivered image quality at 0.25\,arcsec and airmass 1.

Ground-based IFU instruments (such as MUSE, KCWI, or KCRM) might enable a detection and redshift calculation via emission lines even if continuum cannot be reached, but could still require very large amounts of integration time. MUSE, for example, would require on the order of 20 hours of exposure to reach $S/N=5$ on the [O\,{\sc ii}] line of a Kinney Starburst 2 galaxy, shifted to $z=1.2$ and normalised to $R=27.3$. This requires $5\times5$ spatial binning and 20 co-added spectral pixels; however, the [O\,{\sc ii}] doublet would not be resolved at this wavelength binning, making redshift determination dubious.

The point-spread function advantages of a space telescope would increase our chances. WFC3 IR on the \textit{Hubble Space Telescope}\footnote{\url{https://etc.stsci.edu/}}, using the F110W filter, could reach $S/N=5$ on an $R=28.5$ source in one orbit, assuming the Kinney Starburst 2 template. However, the \textit{JWST} is ideal, as a near-infrared instrument, for hunting \scarequotes{high}-redshift objects. Assuming an S0 template normalised to Johnson $R=28.5$, NIRCam with the the F200W filter would reach $S/N\sim120$ in just 33 minutes of 
integration\footnote{\url{https://jwst.etc.stsci.edu/}}.

\subsection{Implications for future FRB hosts}

\autoref{fig:pdfs} and \autoref{fig:zm_all} illustrate that, in the field of \FRBname{}, we could expect to see the majority of FRB hosts well above redshift 1, with the median $z$-$m_\mathrm{R}$ line produced from this sample falling below the statistical limit at redshift 1.8 ($z=2.6$ for \Ks-band). 
Indeed, 10 out of 23 hosts would be seen at $z\sim2$ in $R$-band, and although there is a sharper drop-off at low redshifts, in \Ks-band 9 out of 23 are seen to $z\sim3$. This is encouraging for the prospect of identifying FRB hosts at and beyond cosmic noon; while some hosts will certainly be unseen, some will be luminous enough to detect.  
However, the probability of chance alignment, and hence confusion between the true host and more nearby candidates along the line-of-sight, becomes greater with increasing host redshift.
Conversely, it may occur that a more distant host is identified for an FRB actually emitted from an unseen dim galaxy in the foreground of the identified \scarequotes{host}.
Indeed, \citet{Cordes2022} suggest precisely this case with \FRB{20190611B}, based on a combined \DMFRB{}-\tauFRB{} redshift estimator.
In some ways \FRBname{} is a fortunate case, in which the lack of a host in imaging is quite clear-cut. The potential to not only miss the true host in imaging, but to misidentify another, grows both with the true host's dimness and its cosmological redshift. This will have more ill effects than an obvious non-detection, causing biases in FRB cosmology and in studies of FRB host properties. 

If we conclude that \FRBname{} is at $z>1$, and given that the host of \FRB{20220610A} was detected in 
shallower (with a 5-$\sigma$ depth of $\simeq26$ AB mag)
$R$-band imaging \citep{RyderFRB20220610A}, the implication is that FRB hosts at high redshift occupy a large range of intrinsic luminosities. This is hardly surprising given the rather striking variance in host properties thus far observed in the less-distant Universe, but acts as a reminder to assume little about \scarequotes{typical} FRB hosts at any redshift.

\section{Conclusions}

No host galaxy has been identified for \FRBname{} in spite of deep optical (to $R>\Rmaglim$\,mag) and near-infrared (to $\Ks>\Kmaglim{}$\,mag) follow-up. It is the first instance of a well-localised ASKAP FRB with deep optical follow-up to have not yielded a host galaxy with some certainty, contrasting with \NCRAFTHosts{} 
other ASKAP events (Shannon et al, in preparation). 
Given that it is the first such example in the well-localised CRAFT sample, the lack of detection of a host for \FRBname{} seemed odd at first. However, upon a deeper analysis of the likely redshift of the host and the distribution of FRB host luminosities, we have shown that it is entirely consistent with the range of FRB host galaxies to date, requiring no exotic explanation.
In fact, it may not be at all unusual for higher-redshift FRBs to have unseen hosts.
Further such unseen FRB hosts will accumulate over time. Depending on their prevalence, this may present an obstacle to cosmological applications of FRBs.
If \FRBname{} is at a relatively modest redshift, the problem becomes worse; the proportion of undetected hosts will only grow with distance.

The tension between \FRBname{}'s high fluence ($\FRBfluence$\,Jy\,ms) and the apparent faintness of its host can be resolved by either an unusually dim host or an exceptionally luminous radio burst. Both scenarios are plausible (and even compatible), but suggest that either the burst, its host, or both are outliers in their respective populations.
Barring exotic explanations, the host of \FRBname{} is either intrinsically dim and at $\zhost<0.7$ (Case 2; \autoref{sec:low-z}), or at higher redshift (Case 3; \autoref{sec:high-z}).
If the host of \FRBname{} is more luminous than the R1 (\FRB{20121102A}) host, it must be at $\zhost>0.7$, which would make it the FRB host with at least the second-highest redshift. Confirmation of either would prove scientifically valuable.
We favour Case 3, i.e. that the FRB host is at a relatively large redshift for known FRB hosts, plausibly greater than any known. 
We take this position because: 
 (1) the posterior $p(z|U,\DM{})$ 
 peaks at $z\sim1.2$; 
 (2) ten of the 23 hosts considered here would not be detectable when placed at $z>1$; 
 and 
 (3) 
 the small scattering timescale implies a minor \DMHost{} contribution relative to \DMFRB{}. 
 We predict that the host galaxy of \FRBname{} will eventually be found at $\zhost \approx 1.2$.

Nonetheless, the possibility remains that \FRBname{} could be the first FRB in a host even less luminous than that of R1, itself a significant discovery for the field. 

As more sensitive radio facilities come online, the number of high-redshift FRBs will continue to grow; confirming these redshifts, however, will only become more difficult as they climb higher. While follow-up programmes such as FURBY will continue to associate a majority of ASKAP-detected FRBs with their hosts, the added value of confirming \zhost-\DMFRB{} pairs at high redshift makes those that are missed more bothersome. That a significant proportion of these distant targets could be invisible to all but (or even!) the deepest searches will need to inform the design of future FRB surveys and their follow-up campaigns.

\section*{Acknowledgements}

We thank the referee for their positive and constructive review, which improved the quality of this paper.
C.W.J. and M.G. acknowledge support by the Australian Government through the Australian Research Council's Discovery Projects funding scheme (project DP210102103).
R.M.S. and A.T.D. acknowledge support by the Australian Government through the Australian Research Council's Discovery Projects funding scheme (project DP220102305).
R.M.S. acknowledges support through Australian Research Council Future Fellowship FT190100155.
J.X.P., A.C.G., and N.T. acknowledge support from NSF grants AST-1911140, AST-1910471
and AST-2206490 as members of the Fast and Fortunate for FRB
Follow-up team.
S.B. is supported by a Dutch Research Council (NWO) Veni Fellowship (VI.Veni.212.058). 
H.Q. acknowledges support from Fondation MERAC (Project: SUPERHeRO, PI: Keane)
Parts of this research were supported by the Australian Research Council Centre of Excellence for All Sky Astrophysics in 3 Dimensions (ASTRO 3D), through project number CE170100013.
Based on observations collected at the European Southern Observatory under ESO programmes 105.204W.003 and 108.21ZF.005.
This scientific work uses data obtained from Inyarrimanha Ilgari Bundara, the CSIRO Murchison Radio-astronomy Observatory. We acknowledge the Wajarri Yamaji People as the Traditional Owners and native title holders of the Observatory site. CSIRO’s ASKAP radio telescope is part of the Australia Telescope National Facility (\url{https://ror.org/05qajvd42}). Operation of ASKAP is funded by the Australian Government with support from the National Collaborative Research Infrastructure Strategy. ASKAP uses the resources of the Pawsey Supercomputing Research Centre. Establishment of ASKAP, Inyarrimanha Ilgari Bundara, the CSIRO Murchison Radio-astronomy Observatory and the Pawsey Supercomputing Research Centre are initiatives of the Australian Government, with support from the Government of Western Australia and the Science and Industry Endowment Fund.

%%%%%%%%%%%%%%%%%%%%%%%%%%%%%%%%%%%%%%%%%%%%%%%%%%
\section*{Data Availability}

Code and data for reproducing the results reported here are available at \url{https://github.com/Lachimax/publications/tree/main/Marnoch\%2B2023_FRB20210912A-Missing-Host}.

Optical and infrared imaging data is available from the European Southern Observatory Raw Data Archive (\url{http://archive.eso.org/eso/eso_archive_main.html}).

For other data, please inquire with the lead author.
 
% The inclusion of a Data Availability Statement is a requirement for articles published in MNRAS. Data Availability Statements provide a standardised format for readers to understand the availability of data underlying the research results described in the article. The statement may refer to original data generated in the course of the study or to third-party data analysed in the article. The statement should describe and provide means of access, where possible, by linking to the data or providing the required accession numbers for the relevant databases or DOIs.



\bibliographystyle{mnras}
\bibliography{references_mendeley, references} % if your bibtex file is called example.bib

%\begin{thebibliography}{99}
%\end{thebibliography}

%%%%%%%%%%%%%%%%%%%%%%%%%%%%%%%%%%%%%%%%%%%%%%%%%%

%%%%%%%%%%%%%%%%% APPENDICES %%%%%%%%%%%%%%%%%%%%%

\appendix

\section{PATH Results}
\label{app:path}

We provide, in Tables \ref{tab:g_PATH}, \ref{tab:R_PATH}, \ref{tab:I_PATH} and \ref{tab:K_PATH}, the relevant PATH outputs and inputs for the candidates shown in \autoref{fig:imaging} and summarised in \autoref{tab:path}. ID corresponds to labelling in \autoref{fig:imaging}, $\theta$ is the angular size of the object and $m$ is the apparent magnitude in this image. $P^c$ is the probability of chance coincidence for that object, as described by \citet{PATH} and \citet{Eftekhari17};
$P(O_i)$ is the prior that the object is the host, $p(x|O_i)$ is the marginal likelihood that the object is the host, and $P(O_i|x)$ is the posterior probability that the object is the host, all described in detail by \citet{PATH}.

\begin{table}
\caption{
$g$-band PATH results. 
}
\begin{tabular}{ccccccc}
ID  & $\theta$  & $m$       & $P^c$ & $P(O_i)$ & $p(x|O_i)$ & $P(O_i|x)$ \\
    & (arcsec)  & (AB mag)  &       \\
\\
\hline \\
A & 0.32 & 27.2 & $0.96 \; \mathrm{}$ & $0.047 \; \mathrm{}$ & $1 \times 10^{-40} \; \mathrm{}$ & $9.4 \times 10^{-39} \; \mathrm{}$ \\
C & 0.36 & 27.0 & $0.99 \; \mathrm{}$ & $0.055 \; \mathrm{}$ & $1.7 \times 10^{-80} \; \mathrm{}$ & $1.9 \times 10^{-78} \; \mathrm{}$ \\
D & 0.37 & 26.6 & $0.98 \; \mathrm{}$ & $0.071 \; \mathrm{}$ & $1.3 \times 10^{-76} \; \mathrm{}$ & $1.8 \times 10^{-74} \; \mathrm{}$ \\
E & 0.42 & 25.6 & $0.98 \; \mathrm{}$ & $0.14 \; \mathrm{}$ & $2.5 \times 10^{-192} \; \mathrm{}$ & $6.9 \times 10^{-190} \; \mathrm{}$ \\
F & 0.51 & 26.2 & $1 \; \mathrm{}$ & $0.094 \; \mathrm{}$ & $1.1 \times 10^{-175} \; \mathrm{}$ & $2 \times 10^{-173} \; \mathrm{}$ \\
H & 0.76 & 24.6 & $0.91 \; \mathrm{}$ & $0.3 \; \mathrm{}$ & $6.9 \times 10^{-140} \; \mathrm{}$ & $4.1 \times 10^{-137} \; \mathrm{}$ \\
I & 0.39 & 26.2 & $1 \; \mathrm{}$ & $0.095 \; \mathrm{}$ & $0 \; \mathrm{}$ & $0 \; \mathrm{}$ \\
\end{tabular}
\label{tab:g_PATH}
\end{table}

\begin{table}
\caption{
$R$-band PATH results.
}
\begin{tabular}{ccccccc}
ID  & $\theta$  & $m$       & $P^c$ & $P(O_i)$ & $p(x|O_i)$ & $P(O_i|x)$ \\
    & (arcsec)    & (AB mag)    &       \\
\\
\hline \\
A & 0.47 & 27.0 & $0.94 \; \mathrm{}$ & $0.043 \; \mathrm{}$ & $4.3 \times 10^{-65} \; \mathrm{}$ & $3.7 \times 10^{-63} \; \mathrm{}$ \\
B & 0.30 & 27.6 & $1 \; \mathrm{}$ & $0.028 \; \mathrm{}$ & $0 \; \mathrm{}$ & $0 \; \mathrm{}$ \\
D & 0.44 & 25.6 & $0.87 \; \mathrm{}$ & $0.1 \; \mathrm{}$ & $4.5 \times 10^{-264} \; \mathrm{}$ & $9.1 \times 10^{-262} \; \mathrm{}$ \\
E & 0.55 & 25.0 & $0.9 \; \mathrm{}$ & $0.17 \; \mathrm{}$ & $0 \; \mathrm{}$ & $0 \; \mathrm{}$ \\
F & 0.52 & 26.2 & $1 \; \mathrm{}$ & $0.072 \; \mathrm{}$ & $0 \; \mathrm{}$ & $0 \; \mathrm{}$ \\
H & 0.84 & 24.2 & $0.82 \; \mathrm{}$ & $0.31 \; \mathrm{}$ & $0 \; \mathrm{}$ & $0 \; \mathrm{}$ \\
I & 0.46 & 26.1 & $1 \; \mathrm{}$ & $0.073 \; \mathrm{}$ & $0 \; \mathrm{}$ & $0 \; \mathrm{}$ \\
\end{tabular}
\label{tab:R_PATH}
\end{table}

\begin{table}
\caption{
$I$-band PATH results.
}
\begin{tabular}{ccccccc}
ID  & $\theta$  & $m$       & $P^c$ & $P(O_i)$ & $p(x|O_i)$ & $P(O_i|x)$ \\
    & (arcsec)    & (AB mag)    &       \\
\\
\hline \\
E & 0.28 & 25.5 & $0.98 \; \mathrm{}$ & $0.19 \; \mathrm{}$ & $3.1 \times 10^{-187} \; \mathrm{}$ & $1.2 \times 10^{-184} \; \mathrm{}$ \\
H & 0.44 & 23.9 & $0.75 \; \mathrm{}$ & $0.61 \; \mathrm{}$ & $1.9 \times 10^{-189} \; \mathrm{}$ & $2.3 \times 10^{-186} \; \mathrm{}$ \\
\end{tabular}
\label{tab:I_PATH}
\end{table}

\begin{table}
\caption{
$\Ks$-band PATH results.
}
\begin{tabular}{ccccccc}
ID  & $\theta$  & $m$       & $P^c$ & $P(O_i)$ & $p(x|O_i)$ & $P(O_i|x)$ \\
    & (arcsec)    & (AB mag)    &       \\
\\
\hline \\
G & 0.25 & 24.1 & $0.75 \; \mathrm{}$ & $0.22 \; \mathrm{}$ & $0 \; \mathrm{}$ & $0 \; \mathrm{}$ \\
H & 0.32 & 22.9 & $0.46 \; \mathrm{}$ & $0.58 \; \mathrm{}$ & $0 \; \mathrm{}$ & $0 \; \mathrm{}$ \\
\end{tabular}
\label{tab:K_PATH}
\end{table}

\section{The prior for an unseen host}
\label{app:p_u}

\subsection{The field of FRB~20210912A}

Among the inputs to PATH is a prior that the host is unseen in the image, $P(U)$. 
$P(U)=0$ has typically been assumed for CRAFT host localisations, as in the vast majority of cases deep imaging combined with arcsecond-level positional uncertainty has been sufficient to confidently identify a host. 
However, the assumption is inappropriate for the case of \FRBname{} as it forces PATH to identify, with near-certainty, the closest object in the sky as the host, even if it is too distant to be plausibly so; this is necessitated by the mathematical requirement that the total posterior probability sum to unity.

As noted by \cite{PATH}, the DM of an FRB may be (but thus far has not been) used to inform $P(U)$; as the DM of \FRBname{} (\FRBDM{} $\dmunits$) is high compared to the vast majority of previous precise localisations, we feel safer in adopting a higher $P(U)$, which we set to 0.2. However, this is essentially a guess.

Fortunately, as briefly discussed in \autoref{sec:path}, in the case of this field, this prior has little impact on the end result so long as $P(U) > 0$. When $P(U)$ is incremented from 0 to 0.01 in this field, all values of $P(O_i|x)$ (the posterior probability that object $i$ is the host) fall to a negligible value, in each band, with $P(U|x)$ (the posterior probability that the host is unseen in the imaging) jumping to $\sim1$. Every candidate has a posterior $P(O_i|x) \lll 0.01$ for any $P(U) >= 0.01$; that is, for this field, even a small $P(U)$ goes to a posterior $P(U|x)\sim1$ (differing from unity by, at most, $10^{-34}$). 
This result is consistent in all four imaging bands. 
In $R$-band, with $P(U) = 0.2$, $P(O_i|x)=1\times10^{-65}$ is calculated for Candidate A.
With $P(U) = 0$, it is instead assigned $P(O_i|x)=1$; with $P(U)=0.01$, it is given $P(O_i|x)=3\times10^{-34}$ (and $P(U|x)$ differs from unity by almost the same amount). 

\subsection{A more robust approach}

Although the results of this work are insensitive to $P(U)$ so long as $P(U) > 0$,  this will not be true for many FRB fields. A concrete example of how this prior can significantly affect host identification with PATH is given by \citet{Ibik2023}, comparing results for $P(U)=0.0$ and $P(U)=0.1$ for some CHIME FRBs. The investigation of \FRBname{} has led to some insight into this prior and its use, as a natural extension of the work done here on FRB host luminosities and on the fact of the missing host itself.

The assumption of $P(U)=0$ thus requires rethinking, even in cases where a host galaxy appears obvious by eye; it may artificially inflate the $P(O_i|x)$ value of the most likely host candidate, particularly with borderline cases. It also precludes the possibility of a dwarf host, too faint to be imaged, in chance alignment with or as a satellite of a brighter background candidate. 

In the future regime in which we have large numbers of known (or suspected) FRB hosts, mis-identification has the potential to cause significant bias when conducting analyses of host properties. 
As the number of identifiable hosts grows, it becomes increasingly sensible to generate $P(U)$ algorithmically for consistency and efficiency. Below we outline proposed methods for deriving a more rigorous value, in which information about galaxy populations is combined with a PDF of the source's redshift. The latter can be derived from, as suggested by \cite{PATH}, its DM ($\pzdm$). Additional information about the burst's properties, detector biases and FRB populations may also be accounted for, as demonstrated in \autoref{fig:pdfs} and explained in more detail in \autoref{sec:p_z_dm}.

\subsubsection{Empirical frequentist approach}

The simplest approach to $P(U)$ is to take it as the fraction of other FRBs with an unseen host. Ideally the sample would be partitioned by available imaging depth, band, DM and brightness, but this will require a larger sample of precisely-localised FRBs. 
Even with the entire host sample as input, this method would not be currently feasible, with FRB host statistics still in the small-number regime. It may take a large number of localised FRBs, and non-detections, before $P(U)$ is well-understood empirically. Currently, if we include \FRBname{} and \FRBnameB{} among other well-localised FRBs, it is on the order of 10\%.

\subsubsection{Approach using general galaxy populations}

An algorithm for $P(U)$ might be derived using distributions of galaxy luminosity (`luminosity functions') from the literature (e.g., \citealp{Brown2001, Wolf2003, Willmer2006, Reddy2009, Finkelstein2015}). When these are interpolated (as has been done to produce the $L^*$ lines in \autoref{fig:rmag}) and combined with $P(z|\DM{})$ and an observational magnitude limit derived from the image in question, a value for $P(U)$ can be derived.

An uncertainty in this approach lies in the mapping between galaxy number and FRB probability. Obviously, it makes little sense to attribute every galaxy the same probability of hosting an FRB. However, this depends quite closely on unresolved problems surrounding FRB populations. To first order, it would perhaps make sense to weight FRB probability by stellar mass; however, evidence is mounting that FRBs trace star formation to some degree \citep{GordonProspector, James2022B, JamesH0}. This is less straightforward to map to luminosity, which is the observable of concern here.
While some sort of average star formation-to-luminosity relation would perhaps suffice for this purpose, even this might not be easy to calculate. The picture may be even further complicated by the fact that some small but unknown and perhaps significant fraction of FRBs seem to occur via delayed channels, such as the M81 repeater \citep{GlobularCluster}, offsetting them temporally from star formation. This population could perhaps be approximated to trace stellar mass. Therefore some linear combination of the two scenarios--tracing star formation and tracing stellar mass--would be desirable for maximum precision. As the coefficients for the two scenarios are unknown, this is currently difficult; however, it may be that using the star formation model would suffice to first order.

This method would be complicated, and a new set of luminosity functions would need to be derived for each imaging band (although this becomes less of an issue if PATH is consistently run on $r$-band imaging).

\subsubsection{Approach using an FRB host luminosity function}

The issues mentioned above could perhaps be circumvented by leveraging the high-confidence sample of FRB hosts; instead of concerning ourselves with theoretical mappings between FRB occurrence and galaxy luminosities, we might derive an empirical FRB host luminosity function.
The ideal form of this luminosity function would be one that is allowed to vary with redshift, to account for the evolution of the host population with the Universe; or, more realistically, a set of luminosity functions derived from sets of galaxies partitioned by redshift. However, this is not viable at this point in time; it may become so when hundreds or thousands of FRB hosts are identified, but, for now, host statistics are limited to the population at large.
Even a luminosity function encapsulating the entire host sample is of uncertain merit given the presently small number of known FRB hosts. Nonetheless, it may be attempted. Another weakness to this approach, and the similar approach below, is that previous host misidentifications will bias the result.

\subsubsection{Coarse-grained approach using FRB host luminosities}

Perhaps the simplest available approach is to take our known hosts and calculate their observational magnitudes if placed at a range of plausible redshifts. At any given redshift, 
$P(U|z) = P(\m{} > \m{\mathrm{limit}}|z)$, where we take 
$P(\m{} > \m{\mathrm{limit}}|z) = N_\mathrm{unseen} / N_\mathrm{hosts}$,  simply the fraction of hosts with observational magnitudes above the imaging limit (i.e., in the case of \FRBname{}, the red line in \autoref{fig:pdfs}).
The prior becomes %$P(U) = \sum\limits_n P(U|z_n) P(z_n)$
$P(U) = \int\limits P(U|z) \pzdm \dn z$.

We have tested this approach on the field of \FRBname{}, leveraging the sample of host SED models described in \autoref{sec:sed}, yielding $P(U)=0.10$ in $R$-band, $0.21$ in $g$-band, $0.30$ in $I$-band and $0.08$ in $\Ks$-band.

\subsection{Closing notes}

While $P(U)$ can be informed by DM, as we have explored here, the value of allowing it do so depends on the science goals. 
For instance, FRB cosmology requires independent measurements of \DMFRB\ and \zhost, and if \DMFRB\ is used to inform the host identification they cease to be fully independent. 
Perhaps the suggestions above are more appropriate for cases in which a host association is too uncertain for use with cosmology, 
but could still inform other applications.

Another disadvantage of the population-based methods discussed here is that they require a sample of FRB hosts with SED models. To avoid bias, it would be preferable that this sample be kept updated with each firmly-identified host.
Given the variety that has been repeatedly demonstrated in FRB host properties \citep{Heintz2020, Bhandari2022, GordonProspector}, it is not sufficient to adopt a single template and scale the luminosity to observational magnitudes. Instead, SED fitting of broadband photometry in a number of bands should be undertaken at a minimum, and preferably with the inclusion of spectral information such as performed by \cite{GordonProspector}. This requires extensive follow-up and presents a challenge proportional to the rate of high-precision localisations. 

% If you want to present additional material which would interrupt the flow of the main paper,
% it can be placed in an Appendix which appears after the list of references.

%%%%%%%%%%%%%%%%%%%%%%%%%%%%%%%%%%%%%%%%%%%%%%%%%%

% Don't change these lines
\bsp	% typesetting comment
\label{lastpage}
\end{document}